# Statistical Modelling of the Clipping Noise in OFDM-based Visible Light Communication System

Nima Taherkhani, *Student Member*, *IEEE*, Kamran Kiasaleh, *Senior Member*, *IEEE*

**Abstract**

This paper analyses the statistics of the clipping noise in orthogonal frequency-division-multiplex (OFDM) based visible light Communication systems. The clipped signal is generally modelled as the summation of the scaled original signal and clipping noise, which is treated by the linear equalizer in the receiver. Generally, it is assumed that the clipped and original signal share the same statistics. Although valid in some cases, we show that such assumption is invalid when the transmitter is tightly constrained. We derive closed-form probability distribution function (pdf) for the clipping noise and use the pdf for statistical hypothesis testing in an optimum receiver.

## I.  INTRODUCTION

The emerging 5G networks are predicted to offer high data transmission capacity for new streaming applications. Due to the high definition videos and billions of Internet of Thing (IoT) devices, it is predicted that a capacity per unit area of 100 Mbps will be demanded for future indoor spaces [1]. This goal is getting difficult to achieve for conventional RF communications with limited RF spectra and increasing RF radiation levels in condensed indoor medium. Hence, the alternative wireless transmission technologies have been considered.  VLC has a large free unregulated spectrum, from 428 to 750 THz, which can be exploited for

Nima Taherkhani and Kamran Kiasaleh are with the Erik Jonsson School of Engineering and Computer Science, The University of Texas at Dallas, 800 W. Campbell Road, Richardson, TX, 75080-3021, USA.
(e-mail: nima.taherkhani@utdallas.edu, kamran@utdallas.edu )

connecting the abundant number of IoT devices and for streaming high definition videos [2] . In VLC, the modulating signal is used to switch LEDs at desired frequencies to convey the information. A major limitation in VLC is the low modulation bandwidth of LEDs. To deal with modulation limitation of LED, OFDM scheme has been employed to achieve data rate in the order of gigabits per second (Gbps) [2]. Unlike RF communication, OFDM symbols in VLC need to be real and unipolar. To achieve a real-valued output signal, Hermitian symmetry is applied on the parallel data streams before inverse Fast-Fourier Transform (IFFT) operation. In order to get unipolar transmitting signal, either spectral efficiency or power efficiency needs to be sacrificed in OFDM-based VLC. In DC biased optical OFDM (DCO-OFDM), which results in loss of power efficiency, a bias current is added after multiplexing to make the transmitting signal unipolar [3], while in asymmetrically clipped optical OFDM (ACO-OFDM) [4], where only odd subcarriers carry data, the transmitting signal is clipped at zero. In both of schemes, in order to keep signal transmittable, the transmitting signal is clipped according to the dynamic range of the optical frontend. This clipping will cause nonlinear distortion in the transmitting signal. In a double-side constrained transmitter, where the signal is clipped at the minimum and maximum signal level that an LED can operate, the chance of clipping and its deteriorating impact on transmitting signal increase. This distortion can be severe enough such that it can potentially overtake data carrying and drastically reduces the throughput of the VLC. Similar to the noise in wireless channels, the receiver requires an accurate knowledge about the statistical distribution of the clipping noise in order to battle this distortion and detect and estimate the original data carrying signal. The clipping noise and signal shaping for OFDM is investigated in the literature, where based on the Bussgang theorem, the time-domain clipped Gaussian signal is modelled as an attenuated original signal plus a clipping noise component which is either modelled as a Gaussian random variable [5], or as an uncorrelated noise with unknown distribution [6]. This linear modelling has been used then for deriving the expression for the received symbols in frequency-domain, where the nonlinear clipping distortion is treated as an extra Gaussian noise added to the channel Gaussian noise [7]. Although valid under particular condition, in this paper we question the accuracy of this modelling due to the fact that, for some nontrivial ranges of parameters, Gaussian pdf fails to model the impact of clipping noise, and the exact distribution of this noise in time-domain under practical clipping scenario is yet to be investigated. A key assumption of the previous approach is that the clipped signal remains Gaussian. We demonstrate that the effect of clipping noise imposed by the constraints in VLC may result in a signal statistics that is no longer Gaussian. The motivation to find an alternate distribution is due to the fact that, although the output of the OFDM multiplexer follows a Gaussian distribution due to central limit theorem, the clipping noise resulted by an asymmetrical double-sided clipping will follow an asymmetric distribution. The approximation of the clipping noise, hence, using symmetric distributions, such as Gaussian distribution, will result in an inaccurate modelling of the distortion. With the aid of Kurtosis, we measure the normality of the clipped signal at the output of OFDM multiplexer for non-trivial clipping ranges and show that commonly used linear expression given by Bussgang theory will lead to an accurate answer only if the range specified for truncating the Gaussian random variable is wide enough to leave a better part of the signal intact.

We then theoretically find the distribution function of the clipping noise generated by double-sided clipping of the transmitting signal in the OFDM-based VLC, and use Hellinger distance to validate its accuracy when compared with the empirically-acquired pdf. We also exploit the Kullback-Leibler (K-L) divergence to measure the difference in uncertainty that can be caused by forecasting the clipping noise by a Gaussian distribution and compare it to that of the proposed distribution.

linear expression given by Bussgang theory will lead to an accurate answer only if the range specified for truncating the Gaussian random variable is wide enough to leave a better part of the signal intact.

We then theoretically find the distribution function of the clipping noise generated by double-sided clipping of the transmitting signal in the OFDM-based VLC, and use Hellinger distance to validate its accuracy when compared with the empirically-acquired pdf. We also exploit the Kullback-Leibler (K-L) divergence to measure the difference in uncertainty that can be caused by forecasting the clipping noise by a Gaussian distribution and compare it to that of the proposed distribution.

## II. SYSTEM MODEL OF A DCO-OFDM TRANSMITTER

DCO-OFDM is a form of OFDM that modulates the intensity of an LED, where the clipping and Hermitian symmetry are applied to arrive at a real-valued, non-negative transmitting signal. The block diagram of a DCO-OFDM is presented in Fig.1. The data stream is first mapped onto an M-QAM modulator where the data symbol $S = [S_1, S_2, ..., S_{\frac{N}{2}-1}]$ is generated. To ensure that the output of the multiplexer is a real value signal, the Hermitian symmetry is applied to form the complex symbol vector $I$ given by

$$I = [0, S_1, S_2, ..., S_{\frac{N}{2}-1}, 0, S^*_{\frac{N}{2}-1}, S^*_{\frac{N}{2}-1}, ..., S^*_1]$$

Where $A^*$ denotes the complex conjugate of $A$. IFFT is then applied to $I$ in order to obtain a time domain signal. This results in an N-point IFFT output of the OFDM symbol. For large FFT, *i.e.*, $N > 64$, the output is considered to be Gaussian distributed with zero mean and standard deviation $\sigma_x$, *i.e.*, $x_n \sim N(0, \sigma_x^2)$. Then, the biasing and clipping are performed to transform the FFT output to a unipolar signal. The biasing addition and clipping range are chosen such that the transmitted signal meets the constraints applied by optical power condition and LED dynamic range.

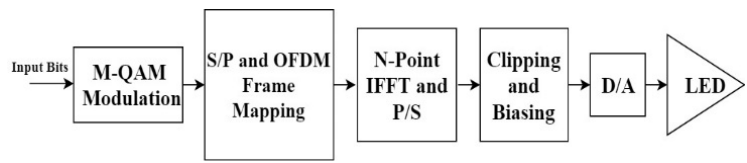

Fig. 1. DCO-OFDM transmitter

For an LED with the dynamic range of $[I_L, I_H]$, the unipolar time domain DCO-OFDM transmitting signal, $x_{DCO,n}$ is given by:

$$x_{DCO,n} = x_{c,n} + I_{bias} \qquad (1)$$

where,

$$x_{c,n} = clip(x_n)$$

$$= \begin{cases} -A_1 & x_n \leq -A_1, \\ x_n & -A_1 < x_n < A_2, \\ A_2 & x_n \geq A_2. \end{cases} \qquad (2)$$

In the above, *clip (.)* denotes the clipping process, $I_{bias}$ is the biasing current, $A_1 = I_{bias} - I_L$ and $A_2 = I_H - I_{bias}$ are the lower and upper clipping bound. In this work, we assume that the minimum operating current of the LED, $I_L$ is zero, which will yield $A_1 = I_{bias}$. The upper and lower clipping ranges are usually set as relative to the standard deviation of the original OFDM symbol $x_n$ using proportionally constant $\alpha_1$ and $\alpha_2$ given by:

$$\begin{aligned} A_1 &= \alpha_1 \sqrt{E\{x_n^2\}} \\ A_2 &= \alpha_2 \sqrt{E\{x_n^2\}} \end{aligned} \qquad (3)$$

where $E\{.\}$ denotes the expectation of the enclosed [8]. The pdf of $x_{c,n}$ can be calculated using (2), and is given by :

$$f(x_{c,n}) =$$

$$\begin{cases} Q(\alpha_1)\delta(x_{c,n} + A_1) & x_{c,n} = -A_1, \\ \frac{1}{\sqrt{2\pi}\sigma_x} e^{-\frac{x_{c,n}^2}{2\sigma_x^2}} & -A_1 < x_{c,n} < A_2, \\ Q(\alpha_2)\delta(x_{c,n} - A_2) & x_{c,n} = A_2. \end{cases} \qquad (4)$$

with $Q(y) = \frac{1}{\sqrt{2\pi}} \int_y^\infty e^{-\frac{v^2}{2}} dv$.

## III. ESTIMATING THE DISTRIBUTION OF CLIPPING NOISE

The clipping process of OFDM signal is generally modeled in the literature using the Bussgang theorem, and is given by [9]:

$$x_{c,n} = \beta x_n + z_n \tag{5}$$

where $\beta$ is the attenuation factor and $z_n$ is the clipping noise, which is assumed to be a random variable uncorrelated with signal $x_n$. Using this property, this factor is calculated by [9]:

$$\beta = \frac{E\{x_{c,n} x_n\}}{E\{x_n^2\}} \tag{6}$$

In this work, we use Kurtosis to analyse the statistical characteristics of (5) and the validity of Gaussian assumption for a desired range of parameters. In the field of statistics, the Kurtosis function is used to measure the peakedness of a pdf and to test the normality of a random variable. The Kurtosis of $x_{c,n}$ can be exploited to measure the normality of the clipped signal as well. In Fig. 2, with the aid simulation, we depict the Kurtosis function of $x_{c,n}$, for different values of clipping bounds, given by :

$$Kurt(x_{c,n}) = \frac{E\{(x_{c,n} - \bar{x}_{c,n})^4\}}{(E\{(x_{c,n} - \bar{x}_{c,n})^2\})^2} \tag{7}$$

A deviation from a Kurtosis of 3 is an indication that the pdf in question is not a Gaussian pdf. As the figure shows, the Kurtosis approaches 3 for only large values of $\alpha_2$ and $\alpha_1$. This is a troubling result as $\alpha_2$ and $\alpha_1$ may assume small values. Hence, we examine the pdf of the clipping noise for a wide range of system parameters. Using (2) and (5), clipping noise can be expressed as a function of $x_n$ for different domains. That is :

$$z_n = g(x_n)$$

$$= \begin{cases} -A_1 - \beta x_n & x_n \leq -A_1, \\ (1-\beta) x_n & -A_1 < x_n < A_2, \\ A_2 - \beta x_n & A_2 \leq x_n. \end{cases} \tag{8}$$

Considering $z_n$ as a function of $x_n$, the cumulative distribution function (cdf) of $z_n$, $n = 1, \ldots, N$, can be calculated using (8). This is given by:

$$F_z(\gamma) = \Pr(z_n < \gamma)$$

$$= \begin{cases} \Pr(x_n > \frac{-A_1-\gamma}{\beta}) & \gamma \geq (1-\beta)A_2 \\ \Pr(\frac{-A_1-\gamma}{\beta} < x_n < \frac{\gamma}{(1-\beta)}) + \Pr(x_n > \frac{A_2-\gamma}{\beta}) & -(1-\beta)A_1 < \gamma < (1-\beta)A_2, \\ \Pr\left(x_n > \frac{A_2-\gamma}{\beta}\right) & \gamma \leq -(1-\beta)A_1. \end{cases} \quad (9)$$

Applying Leibniz Integral rule, the probability density function of $z_n$ is calculated by:

$$f_z(z_n)$$

$$= \begin{cases} \frac{1}{\beta} f_x\left(\frac{-A_1-z_n}{\beta}\right) & z_n \geq (1-\beta)A_2, \\ \frac{1}{(1-\beta)} f_x\left(\frac{z_n}{1-\beta}\right) + \frac{1}{\beta} f_x\left(\frac{-A_1-z_n}{\beta}\right) & \\ + \frac{1}{\beta} f_x\left(\frac{A_2-z_n}{\beta}\right) & -(1-\beta)A_1 < z_n < (1-\beta)A_2, \\ \frac{1}{\beta} f_x\left(\frac{A_2-z_n}{\beta}\right) & z_n \leq -(1-\beta)A_1. \end{cases} \quad (10)$$

where $f_{x_n}(x)$ is the pdf of $x_n$. As the expression in above indicates, $z_n$ follows a Gaussian distribution for a large range of values. However, for a nontrivial domain, the clipping noise follows a mixture distribution which is determined by the summation of three Gaussian pdfs. Although the domain of the mixture distributing is small, it governs the central lobe of $f_z(z_n)$, where the distribution value is much larger than its tail, hence, has a significant importance in approximating the entropy of the clipping noise distribution. It should be also noted that the domain of mixture distribution is basically determined by the clipping bound parameters $\alpha_1$ and $\alpha_2$. For a transmitter with small bounds this domain can get large and, subsequently, make the penalty of Gaussian assumption even more pronounced.

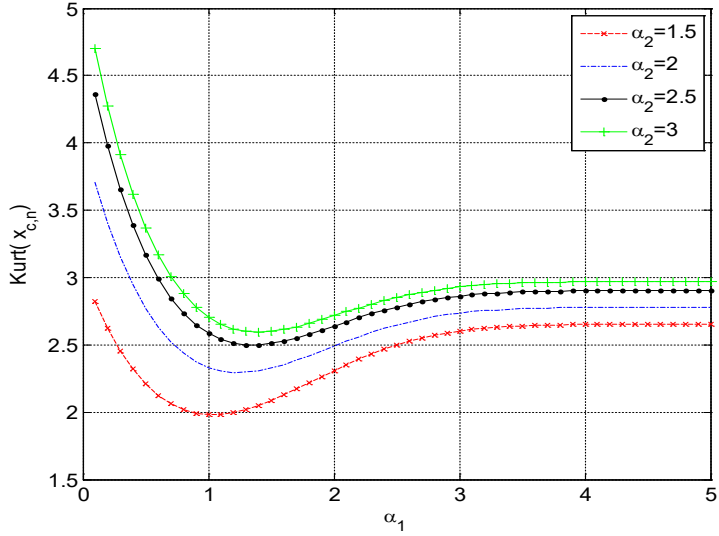

Fig. 2. Kurtosis analysis of double-sided clipped signal $x_{c,n}$.

## IV.     ASSESMENT OF PDF ESTIMATION ACCURACY

In a system whose transmitter has limited dynamic range, the clipping bounds can be fairly small, which will result in a fairly large domain for the mixture distribution. In this section, we use two different metrics, Hellinger distance and Kullback-Leibler (KL) divergence, to quantify the similarity and accuracy of the approximating pdf when compared with simulated pdf of the clipping noise in DCO-OFDM VLC system. Furthermore, the accuracy of the Gaussian pdf in estimating the behaviour of clipping noise is also studied using the aforementioned metrics. In statistics, the Hellinger distance is used to quantify the similarity between two probability distributions, which is defined in terms of the Hellinger integral given by [10] :

$$H(Q, G_i) = \sqrt{1 - \int_{-\infty}^{\infty} \sqrt{q(z)g_i(z)}\,dz} \qquad (11)$$

where $q(z)$ is the empirical pdf obtained by Monte Carlo simulation, which is assumed to estimate the distribution of the clipping noise, and $g_i(z)$ is the distribution under study. We consider two scenarios. That is, $g_1(z) = f_z(z_n)$ and $g_2(z) = \frac{1}{\sqrt{2\pi}\sigma_{ez}} e^{-\frac{(z-\mu_{ez})^2}{2\sigma_{ez}^2}}$, where $\mu_{ez}$ and $\sigma_{ez}$ are the mean and the standard deviation of the Gaussian distribution, which are to be estimated by the maximum likelihood distribution fitting technique of the simulated data. Fig. 3 shows the Hellinger distance between $g_i(z)$ for $i = 1,2$, and the empirical pdf of the clipping noise. As the graph shows, for small clipping bounds $g_1(z)$ significantly outperforms $g_2(z)$ in approximating the underlying distribution. The KL divergence is also used to study how $g_1(z)$ and $g_2(z)$ would diverge from $q(z)$ for different clipping ranges in an optical transmitter. KL divergence for $Q$ and $G_i$, $D_{KL}(Q||G_i)$, is given by [11] :

$$D_{KL}(Q||G_i) = \int_{-\infty}^{\infty} q(z) \log \frac{q(z)}{g_i(z)} dz. \tag{12}$$

This metric can measure the average information lost in approximating $q(z)$ by a Gaussian pdf, *i.e.*, $g_2(z)$.

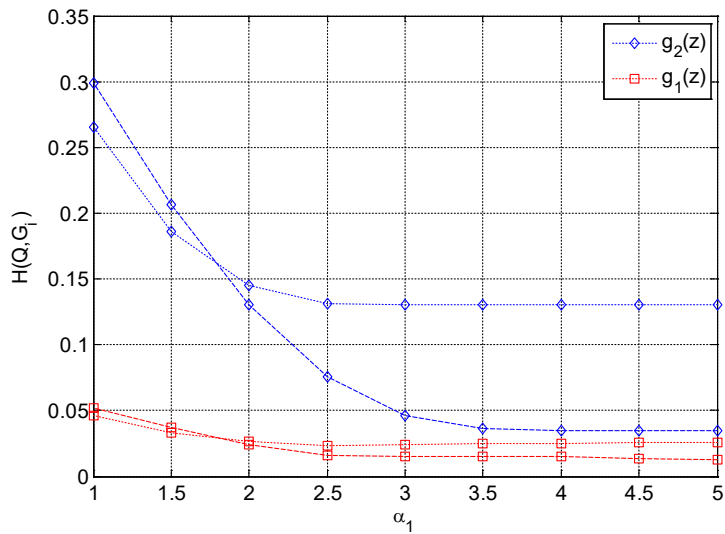

Fig.3: Hellinger distance analysis of $q(z)$ and its two approximating pdfs; *i.e.*, $g_1(z)$ and $g_2(z)$, for different lower clipping bound, when $\alpha_2 = 3$ (dashed lines), $\alpha_2 = 2$ (dotted lines).

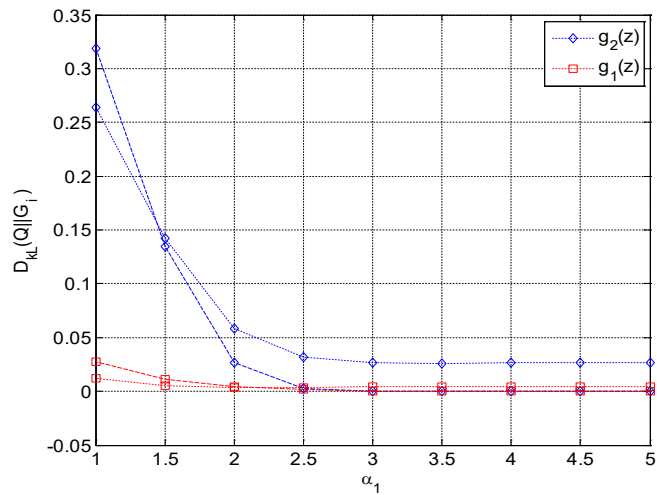

Fig. 4. KL Divergence between $q(z)$ and $g_1(z)$ *vs.* $g_2(z)$, when $\alpha_2 = 3$ (dashed lines), and $\alpha_2 = 2$ (dotted lines).

Fig. 4 depicts the measure of inefficiency and inaccuracy when utilizing a Gaussian distribution as the candidate distribution for the clipping noise. In contrast, the proposed pdf given by (10) yields a fairly small divergence. In a system with a fairly large clipping bound, the Gaussian assumption will yield a KL divergence that is similar to that of the pdf given by (10). However, in a transmitter with a small maximum allowed current, *i.e.*, $\alpha_2 = 2$, the Gaussian pdf diverges considerably from the original distribution for even a large lower clipping bound, *i.e.*, $\alpha_1 = 5$. Shortening of clipping range basically contributes to the spikiness of $q(z)$ and reduces the interval in which it takes non-negligible values. This will make the central lobe of the distribution too sharp and "spikey" and its shoulder too shallow to be modelled by a Gaussian pdf. In contrast, the proposed pdf allows the flexibility of estimating the pdf of the clipping noise in three distinct intervals such that as the clipping bounds change, the margins of these intervals and the clipping attenuation factor $\beta$ in (10) will also change as a function of these bounds, so that even for a system with small bounds the proposed pdf can achieve a close fit .This feature allows one to achieve a high level of accuracy in estimation of the clipping noise for a wide range of system parameters.

## V. CONCLUSION

In this paper, we analysed the statistics of the clipping noise generated in DCO-OFDM VLC due to the physical limitation of the optical front-end. In general, the power constrains that are imposed by the optical frond-end shapes the time domain signal by means of clipping and biasing. This nonlinear modification of transmitting signal leads to a distortion that can be modelled as a Gaussian noise added to the original signal in a linear sense under a particular clipping scenario. However, it was showed in this paper that under stringent clipping conditions, imposed on practical VLC transmitters, this distortion assumes a statistic that deviates from the Gaussian statistics substantially. Subsequently, a new probability distribution function was proposed. In particular, we employed the Bussgang theory to derive a closed-form probability distribution function for the clipping noise as a function of the clipping bounds. The proposed distribution proves to be closely matching the empirical distribution of the clipping noise according to the Hellinger distance and the KL divergence metrics. We note that the design of the corresponding detection strategies for an efficient mitigation of the impact of clipping noise requires an accurate knowledge of its statistical behaviour. The characterization of the clipping noise introduced in this paper will enable system designers to devise more accurate strategies in order to compensate for the impacts of the clipping noise in practical scenarios.